# Gate-modulated reflectance spectroscopy for detecting excitonic species in two-dimensional semiconductors

Mengsong Xue[1,2], Kenji Watanabe[3], Takashi Taniguchi[1], and Ryo Kitaura[1,*]

[1] *Research Center for Materials Nanoarchitectonics (MANA), National Institute for Materials Science (MANA), 1-1 Namiki, Tsukuba 305-0044, Japan*

[2] *Department of Chemistry, Nagoya University, Nagoya 464-8601, Japan*

[3] *Research Center for Functional Materials, National Institute for Materials Science, 1-1 Namiki, Tsukuba 305-0044, Japan*

*Correspondence to KITAURA.Ryo@nims.go.jp

**Abstract**

We have developed a microspectroscopy technique for measuring gate-modulated reflectance to probe excitonic states in two-dimensional transition metal dichalcogenides. Successfully observing excited states of excitons from cryogenic to room temperature showed that this method is more sensitive to excitonic signals than traditional reflectance spectroscopy. Our results demonstrated the potential of this reflectance spectroscopy method in studying exciton physics in two-dimensional transition metal dichalcogenides and their heterostructures.

# Introduction

Various transient quasiparticles, such as excitons and trions, can form in solids in response to light excitation. An exciton, for example, is a bosonic quasiparticle in which a photogenerated electron and a hole form a bound state via attractive Coulomb interaction. Excitons, like phonons, carry momentum and excitation energies, moving in solids before radiative or nonradiative recombination. Electrically neutral excitons can bind an electron or a hole to form charged three-body particles called trions[1–3]. Because of the non-zero net charge, trions can be driven by an electric field, leading to next-generation optoelectronic devices based on electrical manipulations[4–6]. These excitonic quasiparticles usually emerge only at cryogenic temperatures in bulk semiconductors due to weak binding energy between electrons and holes[7–9].

Two-dimensional (2D) semiconductors provide an excellent platform for exploring exciton physics and excitonic devices[10,11]. Due to the reduced dielectric screening arising from 2D structure, exciton binding energies in 2D semiconductors, such as monolayer transition metal dichalcogenides (TMDs), can reach hundreds of millielectronvolts, one to two orders of magnitude higher than in conventional semiconductors[12–14]. For example, the binding energy of monolayer $MoSe_2$ is determined to be 0.55 eV, which is more than 100 times larger than GaAs (4.2 meV)[13,15], enabling the formation of excitons even at room temperature. Due to the colossal binding energy, various excitonic quasiparticles, such as biexcitons[16,17], charged biexciton[18], and five-particle complexes[19,20], are generated in 2D TMDs upon light excitations. Furthermore, moiré excitons emerge in TMD-based moiré superlattices, attracting considerable research attention in the past five years[21–24].

Photoluminescence (PL) spectroscopy has been the primary method to study exciton physics in 2D TMDs. PL spectroscopy is a sensitive probe for low-lying excited states, whereas higher-energy excited states, such as excitonic Rydberg states, are hardly observable[25]. Also, PL spectroscopy detects only radiative recombinations that compete with nonradiative recombinations, and PL intensities can be tiny when nonradiative recombination dominates[26]. Unlike PL spectroscopy, absorption or reflection spectroscopy is independent of the exciton relaxation and recombination processes. As a result, previous studies have successfully observed not only ground states (1s) but also higher-energy excited states (such as 2s) of 2D TMDs[12,25], which were inaccessible by PL spectroscopy, by measuring absorption or reflection contrast (RC).

Here, we have applied an advanced reflectance spectroscopy method, gate-modulated reflectance (GMR) spectroscopy, to probe excitonic states, particularly higher-energy excited states, in 2D TMDs. GMR spectroscopy selectively detects reflectance signals in response to carrier density modulation by AC gate voltage in field-effect transistors (FETs) with semiconductor channels. GMR spectra show an almost flat background compared to standard reflectance spectroscopy because background signal insensitive to carrier density modulation do not contribute to GMR signals. Optical background signals, such as those from polymer residues, are effectively filtered out. As a result, we observed the 2s states of exciton and trion in a monolayer $WS_2$ sample, in which only ground states were observable in standard reflectance spectroscopy at cryogenic temperature. Besides, the 2s state of exciton was observed at room temperature, demonstrating that an exciton, an electron-hole pair, exists even at room temperature. Our work has shown that GMR spectroscopy is a sensitive method to explore exciton physics in 2D TMDs, leading to a further application for investigating exotic excited states, such as moiré excitons in 2D moiré superlattice.

**Results and discussion**

Figures 1(a) and 1(b) show an optical image and a schematic diagram of the device used for GMR measurements. We used an hBN-encapsulated structure to improve the quality of the device, as hBN encapsulation has been shown to effectively narrow excitonic resonances in optical spectra approaching the intrinsic limit[27]. Prior to encapsulation, hBN and WS$_2$ flakes were prepared on Si substrates with a 270 nm SiO$_2$ layer by the mechanical exfoliation method; hBN with a thickness of about 20 nm and monolayer WS$_2$ were identified from optical images. The standard dry transfer technique was used to fabricate an hBN-encapsulated monolayer WS$_2$ on a silicon substrate[28]. Electrical contacts to the monolayer WS$_2$ were made with 20 nm bismuth to maintain low contact resistance down to cryogenic temperatures[29]. As shown below, gate modulation can be applied to a WS$_2$ device over the temperature range of 10 ~ 300 K.

Using the device shown in Fig. 1, we have carried out GMR measurements. Figure 2 shows a diagram of the experimental setup. Monochromatic light with a particular wavelength extracted from a supercontinuum broadband light source was focused on the channel of the device. At the same time, an AC voltage with a specific amplitude and frequency, typically 2 V and 3533 Hz, was applied to the source and drain electrodes for modulating carrier density, while a DC voltage was applied to the Si back gate for controlling average carrier density. Carrier density modulation causes the dielectric function of monolayer TMDs to be modulated, resulting in reflectance modulation, and the lock-in technique was used to selectively detect the reflectance modulation. We made spectra, the modulation amplitude vs. excitation photon energies, by scanning the excitation wavelength with fixed AC amplitudes and frequencies. In this measurement, signals arising from background and impurities such as polymer residues, which are insensitive to the carrier density modulation, do not contribute to the modulation intensity, whereas excitonic peaks, whose oscillator strength, peak position, and peak width strongly depend on carrier density, give significant signals.

Figures 3(a) and 3(b) show RC and GMR spectra of monolayer WS$_2$ measured at 10 K with zero gate voltage, respectively. The RC spectrum was measured with the same setup as the GMR measurement, but the laser power was modulated with a light chopper in the RC measurement instead of carrier density modulation. In our measurements, RC is defined by $(R_{sample} - R_{substrate}) / R_{substrate}$. As shown in Fig. 3(a), no prominent peaks above the noise level were observed in the 2s state energy region (~2.20 eV). Also, in addition to the 1s peaks of excitons ($X^{1s}$) and trions ($T^{1s}$), low-energy peaks (L), possibly due to impurities such as polymer residues, were widely observed in the RC spectrum. On the other hand, the GMR spectrum shows an almost flat background and no low-energy signals which appeared in the RC spectrum.

In contrast to the RC spectrum, we observed additional peaks at the energy region around 0.15 eV higher than 1s peaks in the GMR spectrum (Fig. 3(b)). Due to the interference in multi-layered structures, we analyzed reflectance by transfer-matrix method (TMM)[30]. For analyzing GMR spectra with TMM, the thicknesses of each layer are needed as input parameters. The thicknesses of the top hBN and bottom hBN were obtained by atomic force microscopy (AFM), whereas the thickness of monolayer WS$_2$ was determined based on the interlayer spacing of bulk WS$_2$ (0.618 nm)[31,32]. We also need to know the refractive indices of hBN, Si, and SiO$_2$, as well as the complex dielectric function of WS$_2$ over the spectral range. The dispersion of refractive indices of hBN, Si, and SiO$_2$ was obtained from the literature.[33–35] The complex dielectric function of monolayer WS$_2$, $\varepsilon(E)$, is approximated as a Lorentz oscillator-like model[36]:

$$\epsilon(E) = \epsilon_b + \sum_j \frac{f_j}{E_{0j}^2 - E^2 - i\gamma_j E} \tag{1}$$

where $i$ and $E$ represent the imaginary unit and photon energy, respectively. Also, $\epsilon_b$ represents the background complex dielectric function of monolayer $WS_2$ in the absence of excitons, which is assumed constant and the same as the bulk $WS_2$[37]. The index $j$ represents different excitonic species with oscillator strength $f$, resonant energy $E_0$, and linewidth $\gamma$. In this formalism, each excitonic resonance has three parameters, $f$, $E_0$, and $\gamma$, which can be determined by least-square fitting.

Figures 3(c) and 3(d) show GMR spectra in 1s and 2s energy regions measured at 10 K with a gate voltage of 30 V. Least-square fitting of the 1s energy region reproduces the observed GMR spectra well, yielding resonant energies of 2.072 and 2.041 eV for exciton 1s and trion 1s, respectively. The trion binding energy derived is 31 meV, consistent with PL results[25,38,39]. The peaks in the 2s energy region were also fitted well, assuming the coexistence of 2s exciton ($X^{2s}$) and trion ($T^{2s}$) with resonant energies of 2.215 eV and 2.189 eV, respectively. The energy difference between 1s exciton and 2s exciton is 143 meV, consistent with previous reports[12,25,40]. The binding energy of 2s trion is 26 ± 3 meV, also consistent with a previous report (22 meV)[25]. Using the fitting result, we can calculate the wavelength-dependent absorption coefficient of monolayer $WS_2$ $a(\lambda)$:

$$\alpha(\lambda) = \frac{4\pi k}{\lambda} \tag{2}$$

where $k$ is the imaginary part of the complex refractive index $n(\lambda)$; $n(\lambda) = (\varepsilon(\lambda))^{0.5}$, where $\varepsilon(\lambda)$ represents the complex dielectric function of $WS_2$ as a function of wavelength $\lambda$. The calculated $\lambda$-dependent absorption coefficients, which are identical to the absorption spectra, are presented in the inset, where exciton and trion resonances can be seen as Lorentzian peaks.

Figure 4(a) shows a color plot of a gate voltage dependence of the GMR spectrum in the 2s resonant energy region. As seen in the plot, the GMR signal vanishes below the gate voltage of -20 V due to the depletion of electrons in our naturally n-type $WS_2$ sample. The electron density becomes insensitive to the gate voltage when it drops below the threshold voltage of -20 V, and the GMR signal is no longer present. As electron density increases, the resonant energy of exciton 2s shifts to the higher energy side (Fig. 4(b)). The resonant energy of trion 2s almost keeps constant, indicating the increasing binding energy of trion 2s as electron density increases. This blue shift is consistent with previous reports[41].

Figure 5(a) shows the temperature dependence of the GMR spectra around the 2s-energy region from 10 K to 300 K. As can be seen in the figure, the peaks shift to the lower energy side as the temperature increases, corresponding to the decrease in bandgap; when the temperature is above 50 K, the signature of 2s trion is negligible, which is also consistent with previous reports[25]. We fitted the temperature dependence of exciton 2s peak positions using the Varshni equation:

$$E_g(T) = E_g(0) + \frac{\alpha T^2}{\beta + T}. \tag{3}$$

We fixed $\beta$ to 200 K and obtained $E_g(0)$ = 2.213 eV, $\alpha$ = -3.5 × 10$^{-4}$ eV/K, which is similar to the value in the previous report (-4 × 10$^{-4}$ eV/K)[42]. The fitting results again confirmed that the origin of the high-energy signals is 2s states of excitons. Although peak broadening occurs with increasing temperature, GMR signals from 2s excitons are clearly visible up to room temperature. The existence of the 2s exciton peak at room temperature clearly demonstrates that electron-hole pairs form bound states even at room temperature.

## Conclusion

In conclusion, we have successfully applied GMR spectroscopy to probe excitonic species generated in 2D TMDs. We observed higher-energy GMR signals next to lower-energy signals from exciton and trion ground states; the higher-energy signals were not visible in RC spectra. The TMM-based spectral shape analyses and temperature dependence of peak positions confirmed that these high-energy signals originate from 2s states of exciton and trion. This sensitive and low-noise reflection spectroscopy combined with TMM-based spectral shape analyses is expected to contribute to the study of excitonic physics in 2D semiconductors and their heterostructures.

R.K. was supported by JSPS KAKENHI Grant No. JP23H05469, JP22H05458, JP21K18930 and JP20H05664, and JST CREST Grant No. JPMJCR16F3, SCICORP Grant No. JPMJSC2110 and PRESTO Grant No. JPMJPR20A2. K.W. and T.T. acknowledge support from JSPS KAKENHI Grant Numbers 19H05790, 20H00354, and 21H05233. M.X. was supported by JST SPRING, Grant Number JPMJSP2125. M.X. would like to take this opportunity to thank the "Interdisciplinary Frontier Next-Generation Researcher Program of the Tokai Higher Education and Research System."

# References

[1] M.A. Lampert, "Mobile and Immobile Effective-Mass-Particle Complexes in Nonmetallic Solids," Phys Rev Lett **1**(12), 450–453 (1958).

[2] A. Thilagam, "Two-dimensional charged-exciton complexes," Phys Rev B **55**(12), 7804–7808 (1997).

[3] R.A. Suris, V.P. Kochereshko, G. V Astakhov, D.R. Yakovlev, W. Ossau, J. Nürnberger, W. Faschinger, G. Landwehr, T. Wojtowicz, G. Karczewski, and J. Kossut, "Excitons and Trions Modified by Interaction with a Two-Dimensional Electron Gas," Physica Status Solidi (b) **227**(2), 343–352 (2001).

[4] F. Pulizzi, D. Sanvitto, P.C.M. Christianen, A.J. Shields, S.N. Holmes, M.Y. Simmons, D.A. Ritchie, M. Pepper, and J.C. Maan, "Optical imaging of trion diffusion and drift in GaAs quantum wells," Phys Rev B **68**(20), (2003).

[5] G. Cheng, B. Li, Z. Jin, M. Zhang, and J. Wang, "Observation of Diffusion and Drift of the Negative Trions in Monolayer $WS_2$," Nano Lett **21**(14), 6314–6320 (2021).

[6] T. Hotta, H. Nakajima, S. Chiashi, T. Inoue, S. Maruyama, K. Watanabe, T. Taniguchi, and R. Kitaura, "Trion confinement in monolayer $MoSe_2$ by carbon nanotube local gating," Appl Phys Express **16**(1), (2023).

[7] G.A. Thomas, and T.M. Rice, "Trions, molecules and excitons above the Mott density in Ge," Solid State Commun **23**(6), 359–363 (1977).

[8] T. Kawabata, K. Muro, and S. Narita, "Observation of cyclotron resonance absorptions due to excitonic ion and excitonic molecule ion in silicon," Solid State Commun **23**(4), 267–270 (1977).

[9] B. Stebe, T. Sauder, M. Certier, and C. Comte, "Existence of charged excitons in CuCl," Solid State Commun **26**(10), 637–640 (1978).

[10] G. Wang, A. Chernikov, M.M. Glazov, T.F. Heinz, X. Marie, T. Amand, and B. Urbaszek, "Colloquium: Excitons in atomically thin transition metal dichalcogenides," Rev Mod Phys **90**(2), (2018).

[11] T. Mueller, and E. Malic, "Exciton physics and device application of two-dimensional transition metal dichalcogenide semiconductors," NPJ 2D Mater Appl **2**(1), 29 (2018).

[12] A. Chernikov, T.C. Berkelbach, H.M. Hill, A. Rigosi, Y. Li, O.B. Aslan, D.R. Reichman, M.S. Hybertsen, and T.F. Heinz, "Exciton binding energy and nonhydrogenic Rydberg series in monolayer $WS_2$," Phys Rev Lett **113**(7), (2014).

[13] M.M. Ugeda, A.J. Bradley, S.-F. Shi, F.H. da Jornada, Y. Zhang, D.Y. Qiu, W. Ruan, S.-K. Mo, Z. Hussain, Z.-X. Shen, F. Wang, S.G. Louie, and M.F. Crommie, "Giant bandgap renormalization and excitonic effects in a monolayer transition metal dichalcogenide semiconductor," Nat Mater **13**(12), 1091–1095 (2014).

[14] K. He, N. Kumar, L. Zhao, Z. Wang, K.F. Mak, H. Zhao, and J. Shan, "Tightly Bound Excitons in Monolayer $WSe_2$," Phys Rev Lett **113**(2), 26803 (2014).

[15] S.B. Nam, D.C. Reynolds, C.W. Litton, R.J. Almassy, T.C. Collins, and C.M. Wolfe, "Free-exciton energy spectrum in GaAs," Phys Rev B **13**(2), 761–767 (1976).

[16] T. Hotta, A. Ueda, S. Higuchi, M. Okada, T. Shimizu, T. Kubo, K. Ueno, T. Taniguchi, K. Watanabe, and R. Kitaura, "Enhanced Exciton-Exciton Collisions in an Ultraflat Monolayer $MoSe_2$ Prepared through Deterministic Flattening," ACS Nano **15**(1), 1370–1377 (2021).

[17] M. Okada, Y. Miyauchi, K. Matsuda, T. Taniguchi, K. Watanabe, H. Shinohara, and R. Kitaura, "Observation of biexcitonic emission at extremely low power density in tungsten disulfide atomic layers grown on hexagonal boron nitride," Sci Rep **7**(1), (2017).

[18] Z. Ye, L. Waldecker, E.Y. Ma, D. Rhodes, A. Antony, B. Kim, X.X. Zhang, M. Deng, Y. Jiang, Z. Lu, D. Smirnov, K. Watanabe, T. Taniguchi, J. Hone, and T.F. Heinz, "Efficient generation of neutral and charged biexcitons in encapsulated $WSe_2$ monolayers," Nat Commun **9**(1), (2018).

[19] S.Y. Chen, T. Goldstein, T. Taniguchi, K. Watanabe, and J. Yan, "Coulomb-bound four- and five-particle intervalley states in an atomically-thin semiconductor," Nat Commun **9**(1), (2018).

[20] Z. Li, T. Wang, Z. Lu, C. Jin, Y. Chen, Y. Meng, Z. Lian, T. Taniguchi, K. Watanabe, S. Zhang, D. Smirnov, and S.F. Shi, "Revealing the biexciton and trion-exciton complexes in BN encapsulated $WSe_2$," Nat Commun **9**(1), (2018).

[21] K.L. Seyler, P. Rivera, H. Yu, N.P. Wilson, E.L. Ray, D.G. Mandrus, J. Yan, W. Yao, and X. Xu, "Signatures of moiré-trapped valley excitons in $MoSe_2/WSe_2$ heterobilayers," Nature **567**(7746), 66–70 (2019).

[22] E.M. Alexeev, D.A. Ruiz-Tijerina, M. Danovich, M.J. Hamer, D.J. Terry, P.K. Nayak, S. Ahn, S. Pak, J. Lee, J.I. Sohn, M.R. Molas, M. Koperski, K. Watanabe, T. Taniguchi, K.S. Novoselov, R. V. Gorbachev, H.S. Shin, V.I. Fal'ko, and A.I. Tartakovskii, "Resonantly hybridized excitons in moiré superlattices in van der Waals heterostructures," Nature **567**(7746), 81–86 (2019).

[23] K. Tran, G. Moody, F. Wu, X. Lu, J. Choi, K. Kim, A. Rai, D.A. Sanchez, J. Quan, A. Singh, J. Embley, A. Zepeda, M. Campbell, T. Autry, T. Taniguchi, K. Watanabe, N. Lu, S.K. Banerjee, K.L. Silverman, S. Kim, E. Tutuc, L. Yang, A.H. MacDonald, and X. Li, "Evidence for moiré excitons in van der Waals heterostructures," Nature **567**(7746), 71–75 (2019).

[24] C. Jin, E.C. Regan, A. Yan, M. Iqbal Bakti Utama, D. Wang, S. Zhao, Y. Qin, S. Yang, Z. Zheng, S. Shi, K. Watanabe, T. Taniguchi, S. Tongay, A. Zettl, and F. Wang, "Observation of moiré excitons in $WSe_2/WS_2$ heterostructure superlattices," Nature **567**(7746), 76–80 (2019).

[25] A. Arora, T. Deilmann, T. Reichenauer, J. Kern, S. Michaelis De Vasconcellos, M. Rohlfing, and R. Bratschitsch, "Excited-State Trions in Monolayer $WS_2$," Phys Rev Lett **123**(16), (2019).

[26] M. Amani, D.-H. Lien, D. Kiriya, J. Xiao, A. Azcatl, J. Noh, S.R. Madhvapathy, R. Addou, S. Kc, and M. Dubey, "Near-unity photoluminescence quantum yield in $MoS_2$," Science (1979) **350**(6264), 1065–1068 (2015).

[27] F. Cadiz, E. Courtade, C. Robert, G. Wang, Y. Shen, H. Cai, T. Taniguchi, K. Watanabe, H. Carrere, D. Lagarde, M. Manca, T. Amand, P. Renucci, S. Tongay, X. Marie, and B. Urbaszek, "Excitonic linewidth approaching the homogeneous limit in $MoS_2$-based van der Waals heterostructures," Phys Rev X **7**(2), (2017).

[28] S. Masubuchi, M. Morimoto, S. Morikawa, M. Onodera, Y. Asakawa, K. Watanabe, T. Taniguchi, and T. Machida, "Autonomous robotic searching and assembly of two-dimensional crystals to build van der Waals superlattices," Nat Commun **9**(1), 1413 (2018).

[29] P.C. Shen, C. Su, Y. Lin, A.S. Chou, C.C. Cheng, J.H. Park, M.H. Chiu, A.Y. Lu, H.L. Tang, M.M. Tavakoli, G. Pitner, X. Ji, Z. Cai, N. Mao, J. Wang, V. Tung, J. Li, J. Bokor, A. Zettl, C.I. Wu, T. Palacios, L.J. Li, and J. Kong, "Ultralow contact resistance between semimetal and monolayer semiconductors," Nature **593**(7858), 211–217 (2021).

[30] S.J. Byrnes, "Multilayer optical calculations," ArXiv:1603.02720, (2016).

[31] J.A. Wilson, and A.D. Yoffe, "The transition metal dichalcogenides discussion and interpretation of the observed optical, electrical and structural properties," Adv Phys **18**(73), 193–335 (1969).

[32] Y. Li, A. Chernikov, X. Zhang, A. Rigosi, H.M. Hill, A.M. van der Zande, D.A. Chenet, E.M. Shih, J. Hone, and T.F. Heinz, "Measurement of the optical dielectric function of monolayer transition-metal dichalcogenides: $MoS_2$, $MoSe_2$, $WS_2$, and $WSe_2$," Phys Rev B **90**(20), (2014).

[33] Y. Rah, Y. Jin, S. Kim, and K. Yu, "Optical analysis of the refractive index and birefringence of hexagonal boron nitride from the visible to near-infrared," Opt Lett **44**(15), 3797 (2019).

[34] M.A. Green, "Self-consistent optical parameters of intrinsic silicon at 300 K including temperature coefficients," Solar Energy Materials and Solar Cells **92**(11), 1305–1310 (2008).

[35] I.H. Malitson, "Interspecimen Comparison of the Refractive Index of Fused Silica," J Opt Soc Am **55**(10), 1205–1209 (1965).

[36] A. Arora, A. Mandal, S. Chakrabarti, and S. Ghosh, "Magneto-optical Kerr effect spectroscopy based study of Landé g-factor for holes in GaAs/AlGaAs single quantum wells under low magnetic fields," J Appl Phys **113**(21), (2013).

[37] H. Wang, G. Qin, G. Li, H. Kang, H.S. Yu, J. Baik, A.R. Beal, Y. Liang, and H.P. Hughes, "Kramers-Kronig Analysis of the Reflectivity Spectra of $3R-WS_2$ and $2H-WSe_2$," J. Phys. C: Solid State Phys. **9** 2449 (1976).

[38] M. Zinkiewicz, A.O. Slobodeniuk, T. Kazimierczuk, P. Kapuściński, K. Oreszczuk, M. Grzeszczyk, M. Bartos, K. Nogajewski, K. Watanabe, T. Taniguchi, C. Faugeras, P. Kossacki, M. Potemski, A. Babiński, and M.R. Molas, "Neutral and charged dark excitons in monolayer $WS_2$," Nanoscale **12**(35), 18153–18159 (2020).

[39] M. Zinkiewicz, T. Woźniak, T. Kazimierczuk, P. Kapuscinski, K. Oreszczuk, M. Grzeszczyk, M. Bartoš, K. Nogajewski, K. Watanabe, T. Taniguchi, C. Faugeras, P. Kossacki, M. Potemski, A. Babiński, and M. R. Molas, "Excitonic Complexes in n-Doped $WS_2$ Monolayer," Nano Lett **21**(6), 2519–2525 (2021).

[40] H.M. Hill, A.F. Rigosi, C. Roquelet, A. Chernikov, T.C. Berkelbach, D.R. Reichman, M.S. Hybertsen, L.E. Brus, and T.F. Heinz, "Observation of excitonic rydberg states in monolayer $MoS_2$ and $WS_2$ by photoluminescence excitation spectroscopy," Nano Lett **15**(5), 2992–2997 (2015).

[41] K. Wagner, E. Wietek, J.D. Ziegler, M.A. Semina, T. Taniguchi, K. Watanabe, J. Zipfel, M.M. Glazov, and A. Chernikov, "Autoionization and Dressing of Excited Excitons by Free Carriers in Monolayer $WSe_2$," Phys Rev Lett **125**(26), (2020).

[42] G. Plechinger, P. Nagler, J. Kraus, N. Paradiso, C. Strunk, C. Schüller, and T. Korn, "Identification of excitons, trions and biexcitons in single-layer $WS_2$," Phys Status Solidi RRL **9**(8), 457–461 (2015).

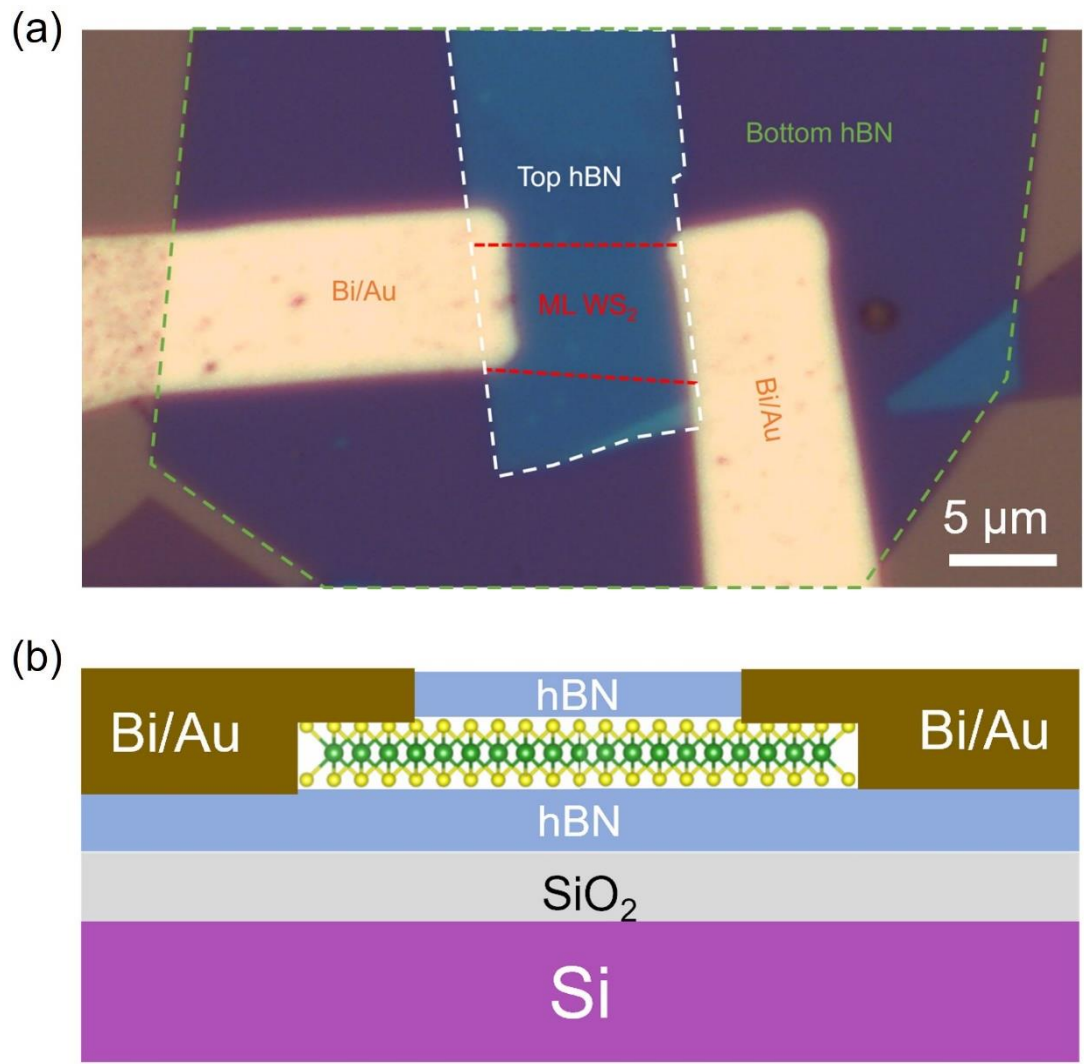

Figure 1. (a) Optical image of the two-terminal device of the hBN/WS$_2$/hBN heterostructure. (b) Schematic representation of a cross-section of the device. Bi directly touches the WS$_2$ flake to make contact resistant small.

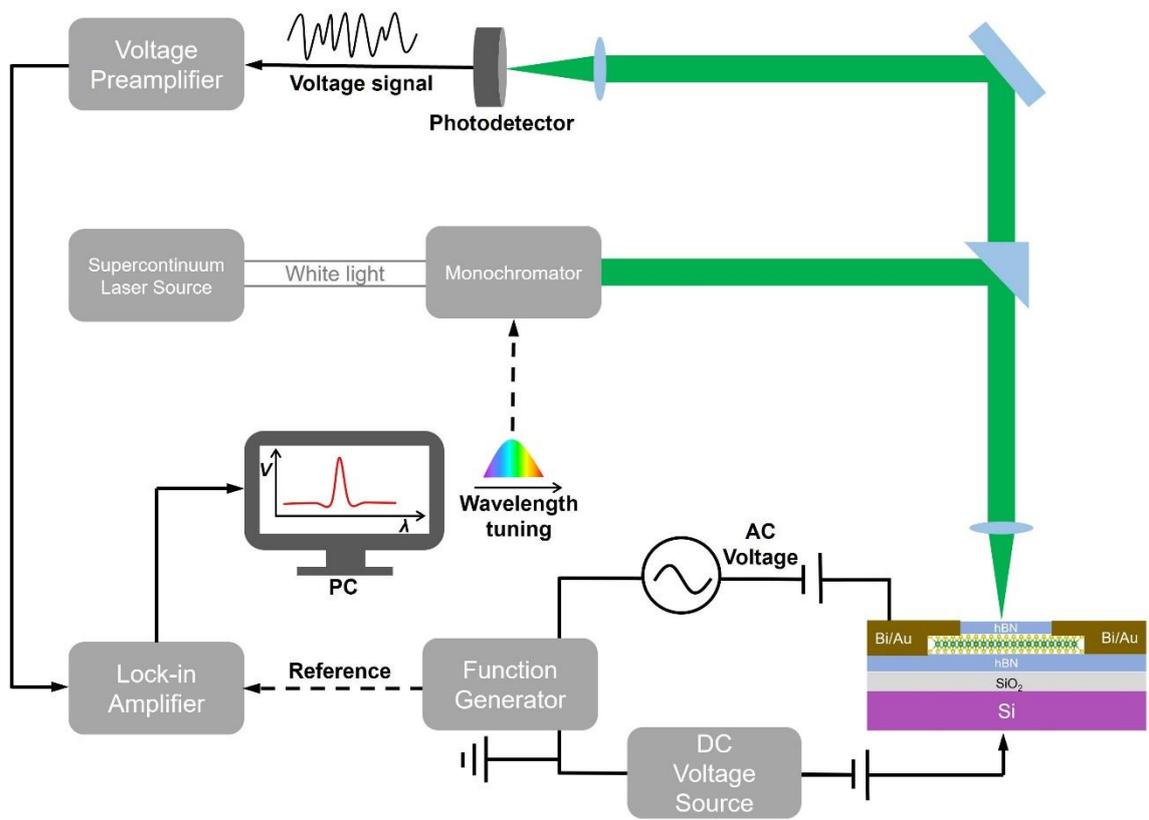

Figure 2. Schematic representation of the setup for gate-modulated reflectance spectroscopy.

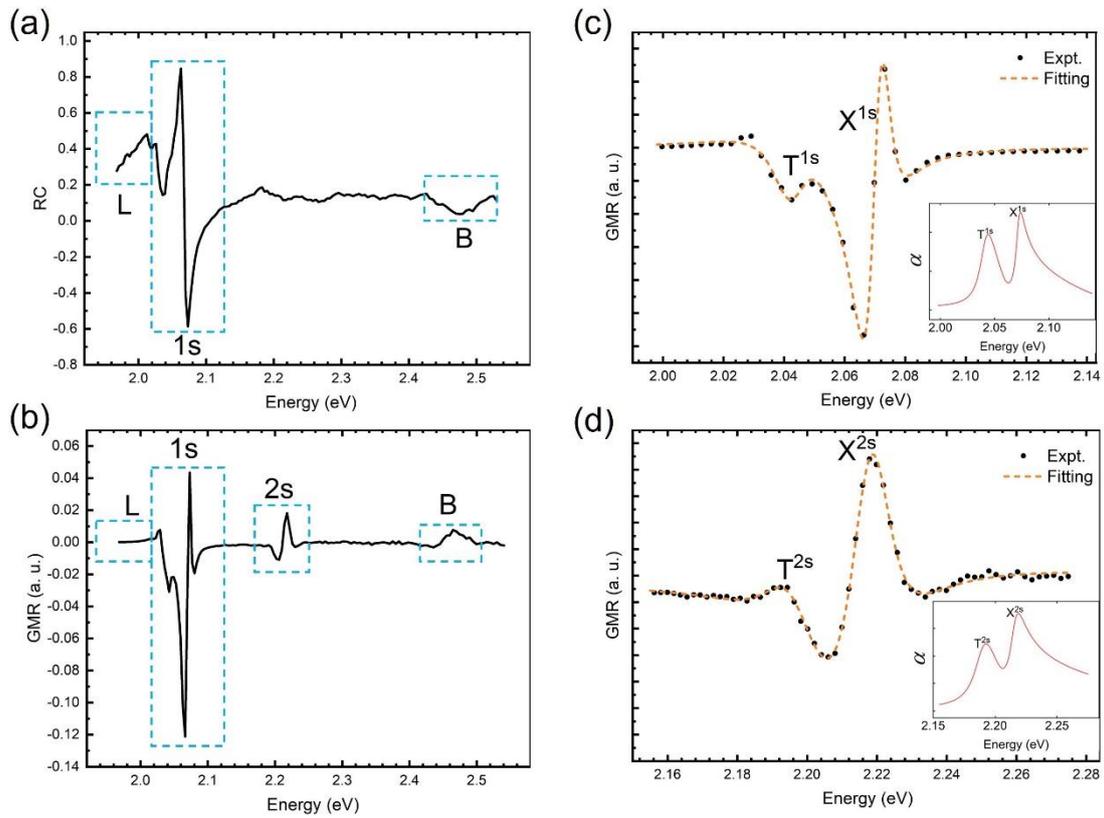

Figure. 3. (a, b) RC spectrum and GMR spectrum of the sample measured at 10 K. (c, d) GMR spectra of the 1s (c) and 2s (d) energy regions. Black dots are the experimental data and orange dashed lines are the fits with the TMM method. The insets show the calculated absorption coefficient of monolayer $WS_2$.

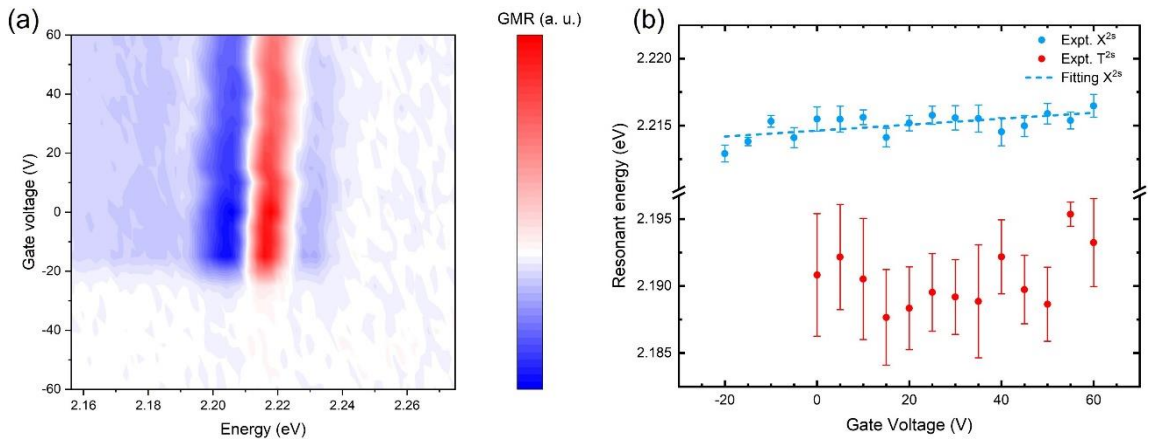

Figure. 4. (a) Color plot of the gate-dependent GMR spectra. (b) Plot of the gate dependence of exciton 2s (blue) resonant energies and trion 2s (red) states with error bars. Dashed line is the linear fit.

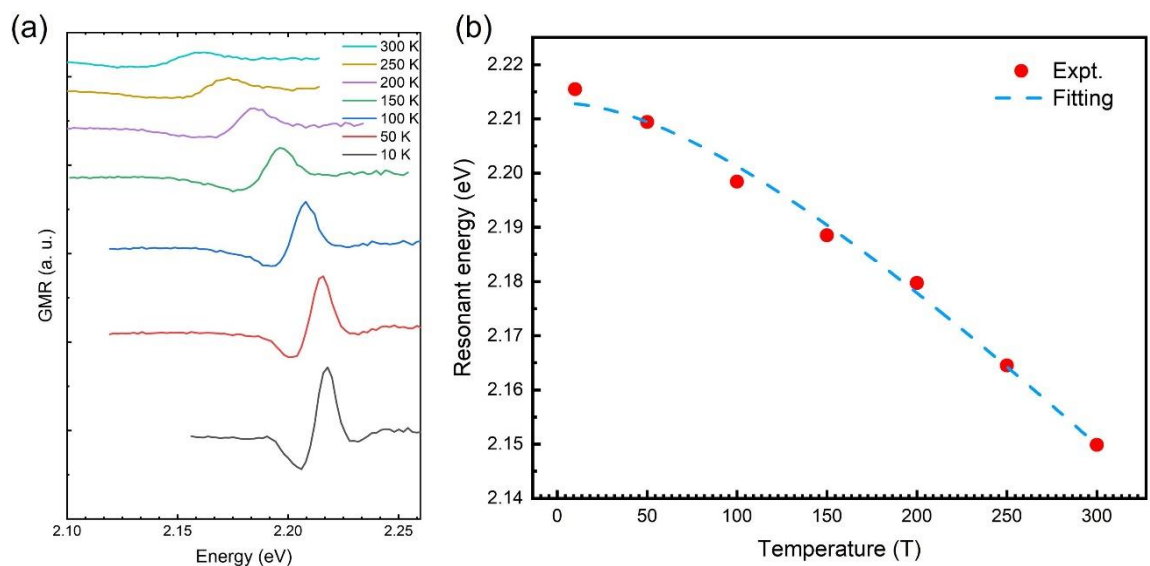

Figure. 5. (a) Temperature-dependent GMR spectra of the sample measured from 10 K to 300 K. (b) Temperature dependence of the resonant energy of exciton 2s state extracted from (a). Red dots are experimental results and the blue dashed line is a fitting curve with the Varshni equation.